\documentstyle[12pt]{article}
\renewcommand{\theequation}{\arabic{equation}}
\newcommand{\EQ}{\begin{equation}}
\newcommand{\EN}{\end{equation}}

\newcommand{\ket}[1]{\left|#1\right\rangle}      

\newcommand{\bear}{\begin{eqnarray}}
\newcommand{\ear}{\end{eqnarray}}
\newcommand{\bt} { \begin{tabular} }
\newcommand{\et}{ \end{tabular} }
\newcommand{\bc} { \begin{center} }
\newcommand{\ec}{ \end{center} }

\newcommand{\btb} { \begin{table} }
\newcommand{\etb}{ \end{table} }

\begin{document}

\topmargin 0pt
\oddsidemargin 5mm
\newcommand{\NP}[1]{Nucl.\ Phys.\ {\bf #1}}
\newcommand{\PL}[1]{Phys.\ Lett.\ {\bf #1}}
\newcommand{\NC}[1]{Nuovo Cimento {\bf #1}}
\newcommand{\CMP}[1]{Comm.\ Math.\ Phys.\ {\bf #1}}
\newcommand{\PR}[1]{Phys.\ Rev.\ {\bf #1}}
\newcommand{\PRL}[1]{Phys.\ Rev.\ Lett.\ {\bf #1}}
\newcommand{\MPL}[1]{Mod.\ Phys.\ Lett.\ {\bf #1}}
\newcommand{\JETP}[1]{Sov.\ Phys.\ JETP {\bf #1}}
\newcommand{\TMP}[1]{Teor.\ Mat.\ Fiz.\ {\bf #1}}

\renewcommand{\thefootnote}{\fnsymbol{footnote}}

\newpage
\setcounter{page}{0}
\begin{titlepage}
\begin{flushright}
UFSCARF-TH-03-14
\end{flushright}
\vspace{0.5cm}
\begin{center}
{\large Integrable $SU(N)$ vertex models with general toroidal boundary conditions}\\
\vspace{1cm}
{\large G.A.P. Ribeiro, M.J. Martins and W. Galleas} \\
\vspace{1cm}
{\em Universidade Federal de S\~ao Carlos\\
Departamento de F\'{\i}sica \\
C.P. 676, 13565-905~~S\~ao Carlos(SP), Brasil}\\
\end{center}
\vspace{0.5cm}

\begin{abstract}
We formulate the algebraic Bethe ansatz solution of the $SU(N)$ vertex models with rather
general non-diagonal toroidal boundary conditions.  The reference states needed in the Bethe
ansatz construction are found by performing gauge transformations 
on the Boltzmann weights in the manner of Baxter \cite{BA}.
The structure of the transfer matrix eigenvectors consists of multi-particle
states over such pseudovacuums and the corresponding eigenvalues depend crucially on the
boundary matrix eigenvalues. We also discuss for $N=2$ the peculiar case of twisted 
boundaries associated to singular matrices.

\end{abstract}

\vspace{.15cm}
\centerline{PACS numbers:  05.50+q, 02.30.IK}
\vspace{.1cm}
\centerline{Keywords: Algebraic Bethe Ansatz, Lattice Models}
\vspace{.15cm}
\centerline{July 2003}
\end{titlepage}

\renewcommand{\thefootnote}{\arabic{footnote}}

\section{Introduction}

The study of vertex models have led to important developments in the field of exactly solvable
models in two dimensions \cite{BA}. Their transfer matrices are in general constructed from local
Boltzmann weights ${\cal L}_{{\cal A}i}(\lambda)$ where $\lambda$ denotes a spectral parameter. This
operator can be viewed as a matrix on the space of states $\cal A$ representing, for instance, the horizontal
degrees of freedom of the vertex model on the square lattice. Its matrix elements are operators 
on $\displaystyle \prod_{i=1}^{L} \otimes V_{i} $
where $V_{i}$ represents the space of states of the vertical degrees of freedom at each site $i$ of a chain of size $L$. The
corresponding transfer matrix can be expressed in terms of an ordered product of ${\cal L}_{{\cal A}i}(\lambda)$ operator over the space ${\cal A}$
denominated monodromy operator ${\cal T}_{{\cal A}}(\lambda)$  \cite{FA},
\EQ
{\cal T}_{{\cal A}}(\lambda)={\cal L}_{{\cal A}L}(\lambda){\cal L}_{{\cal A}L-1}(\lambda) \dots {\cal L}_{{\cal A}1}(\lambda)
\EN

In terms of the monodromy matrix, a sufficient condition for integrability is the Yang-Baxter algebra \cite{FA,KO} which reads
\EQ
R(\lambda, \mu) {\cal T}_{{\cal A}}(\lambda) \otimes {\cal T}_{{\cal A}}(\mu) =
{\cal T}_{{\cal A}}(\mu) \otimes {\cal T}_{{\cal A}}(\lambda) R(\lambda, \mu)
\label{fundrel}
\EN
where $R(\lambda,\mu)$ is an invertible matrix over complex numbers acting on the tensor product ${\cal A} \otimes {\cal A}$ space.

The Yang-Baxter algebra is invariant by the transformation ${\cal T}_{{\cal A}}(\lambda) \rightarrow {\cal G}_{\cal A} {\cal T}_{{\cal A}}(\lambda)$
provided that the group of c-numbers matrices ${\cal G}_{\cal A}$ satisfies the following property \cite{VE}
\EQ
[R(\lambda, \mu), {\cal G}_{\cal A} \otimes {\cal G}_{\cal A}]=0\label{Gcondition}
\EN

An immediate consequence of this symmetry is the possibility to define the operator
\EQ
T(\lambda)=Tr_{{\cal A}}[{ {\cal G}_{\cal A} \cal T_{\cal A}(\lambda)}]\label{transferT1}
\EN
which gives origin to generalized families of commuting transfer matrices.

When the matrix ${\cal G}_{{\cal A}}$ is non singular a quantum spin chain can be associated with the transfer matrix (\ref{transferT1}).
For sake of simplicity, consider the usual situation in which the spaces ${\cal A}$ and $V_{i}$ are isomorphic and that the
${\cal L}_{{\cal A}i}(\lambda)$ is proportional to the exchange operator $P_{{\cal A}i}$  at  certain special point say $\lambda=0$. The
corresponding one-dimensional Hamiltonian is obtained as a 
logarithmic derivative of the transfer matrix at point $\lambda=0$, which reads \cite{VE,BAT1}
\EQ
{\cal H}=\sum_{i=1}^{L-1} P_{ii+1}\frac{d{\cal L}_{ii+1}(\lambda)}{d\lambda}\mid_{\lambda=0}+{\cal G}_{L}^{-1}P_{L 1}\frac{d{\cal L}_{L 1}(\lambda)}{d\lambda}\mid_{\lambda=0}{\cal G}_{L}
\label{HAMIL}
\EN

Clearly, the admissible ${\cal G}_{\cal A}$ matrices play the role of more general toroidal boundary conditions than the particular periodic
case when ${\cal G}_{\cal A}$ is the identity matrix, the simplest possibility satisfying relation (\ref{Gcondition}). From the point of
view of a vertex model, such general twisted boundary conditions correspond to the introduction of a line of defects along the infinite
direction on the cylinder. Though boundary conditions are not expected to influence the infinite volume properties it can change the finite-size corrections
of massless systems in a strip of width $L$ which contains fundamental informations concerning the underlying conformal field theories \cite{CA}.
For instance, in statistical mechanics boundary conditions provide the means to relate the critical behaviour of a variety of different
lattice systems such as the Heisenberg spin chain, the Ashkin-Teller and the Potts models \cite{AL}. In this
sense, it is highly desirable to study integrable models with as much general boundary conditions as possible.

If the boundary matrix ${\cal G}_{\cal A}$ is diagonal the corresponding transfer matrix (\ref{transferT1})  
can be diagonalized with very
little difference from the periodic case because it does not change in a drastic way the properties of the monodromy matrix
elements. The same does not occur when ${\cal G}_{\cal A}$ is non-diagonal, starting from the fact that the reference state
of the periodic case, essential to implement Bethe ansatz approaches, is a 
priori no longer of utility due to the breaking of the original
bulk symmetry by the boundary terms. In fact, progress towards 
solving commuting transfer matrices with general twists by Bethe ansatz techniques 
are modest as compared with the literature known for the periodic case, specially for solvable vertex
models based on Lie algebras, e.g. refs.\cite{RE,KU,MA}. To our knowledge, the six vertex 
model and its higher spin descendants \cite{BAT,BAT1} are
the only solvable vertex systems analyzed so far with non-diagonal boundary conditions. 
Even in these cases, the functional relation method used in 
refs.\cite{BAT,BAT1} gives the transfer matrix eigenvalues 
but not information on
the corresponding eigenvectors. The latter is certainly an important step in the program of solving integrable
systems.

The purpose of this paper is the formulation 
of the quantum inverse scattering method for the simplest multistate generalization
of the six vertex model having $N$ independent degrees of freedom on 
each lattice bond. This turns out to be the isotropic $SU(N)$ vertex
model whose origin goes back to the work by Uimin \cite{UI} and  Sutherland \cite{SU} on 
generalized integrable Heisenberg chains with higher symmetry.
Its corresponding ${\cal L}_{{\cal A}i}(\lambda)$ operators can be written as \cite{KUL,DE}
\EQ
{\cal L}_{{\cal A}i}(\lambda) = \lambda I_{{\cal A}i} + P_{{\cal A}i}
\EN
where as usual $I_{{\cal A}i}$ is the identity matrix on the space ${\cal A} \otimes V_{i}$.

The interesting feature of this system is that the admissible 
symmetries constitute of arbitrary $N \times N$ ${\cal G}_{\cal A}$ matrices due to the standard property
$P_{12} A_1 \otimes B_2 = B_2 \otimes A_1 P_{12}$. Therefore this provides  
us a rich variety of possible diagonal and non-diagonal boundary conditions.
In next section, we present the details of the solution of the eigenvalue problem for the transfer matrix 
in the simplest $N=2$ case. Interesting enough, we
find that the Bethe 
ansatz solution depends on the eigenvalue problem related to the boundary ${\cal G}_{\cal A}$
matrix. In section \ref{Nested Bethe ansatz for SUN model} we generalize these 
results for arbitrary values of $N$ by using the nested Bethe
ansatz approach. Our conclusions are presented 
in section \ref{Conclusion} as well as a discussion of singular boundaries for the model $N=2$. 
In Appendix A we discuss briefly the completeness of the Hilbert space for $N=2$ and finite $L$.

\section{Algebraic Bethe ansatz for Heisenberg model}\label{Algebraic Bethe ansatz for SU2 model}

The purpose of this section is to determine the eigenvalues and the eigenvectors of the following transfer matrix
\EQ
T(\lambda)=Tr_{\cal A}[{ {\cal G}_{\cal A} {\cal L}_{{\cal A} L}(\lambda)\dots {\cal L}_{{\cal A} 1}(\lambda) }]
\label{transferT}
\EN

The operator ${\cal L}_{{\cal A} i}(\lambda)$ is the elementary Boltzmann weights of the isotropic six vertex model which can be written as
\EQ
{\cal L}_{{\cal A} i}(\lambda)=\left( \begin{array}{cc}
        \frac{[a(\lambda)+b(\lambda)]}{2} I_{i}+ \frac{[a(\lambda)-b(\lambda)]}{2}\sigma_{i}^{z} &   \sigma_{i}^{-} \\
        \sigma_{i}^{+} &  \frac{[a(\lambda)+b(\lambda)]}{2} I_{i}-\frac{[a(\lambda)-b(\lambda)]}{2} \sigma_{i}^{z}  \\
        \end{array}\right)
\EN
where $\sigma_{{\cal A}, i}^{\pm}$ and $\sigma_{{\cal A}, i}^{z}$ are Pauli matrices acting on the vertical space of states and the weights are
$a(\lambda)=\lambda+1$ and $b(\lambda)=\lambda$. The boundary matrix ${\cal G}_{{\cal A}}$ is an arbitrary $2 \times 2$ matrix over
the complex numbers whose matrix elements are denoted by
\EQ
{\cal G}_{\cal A}=\left( \begin{array}{cc}
        g_{11} &   g_{12} \\
        g_{21} &   g_{22} \\
        \end{array}\right)
\EN

An essential ingredient of the quantum inverse scattering is the existence of 
a reference state such that the action of the monodromy operator in this state gives as a result a
triangular matrix. Though each of the operators 
${\cal L}_{{\cal A} i}(\lambda)$ when acting on the trivial spin up $\left(\begin{array}{c} 1 \\ 0 \\ \end{array}\right)_{i}$ or spin down
$\left(\begin{array}{c} 0 \\ 1 \\ \end{array}\right)_{i}$ states becomes triangular, such property is not extended to the total
monodromy because the off diagonal elements of ${\cal G}_{\cal A}$ are in general non-null. Therefore the standard ferromagnetic pseudovacuum
is not useful when both $g_{12}$ and $g_{21}$ are different from zero. In order to find an appropriate reference state we have to introduce
a set of gauge transformations similar to that used by Baxter \cite{BA} in the solution of the eight vertex model. We replace the local operators
${\cal L}_{{\cal A} i}(\lambda)$ by new matrices $\widetilde{{\cal L}}_{{\cal A} j}(\lambda)$ such that \cite{FA}
\EQ
\widetilde{{\cal L}}_{{\cal A} j}(\lambda)=M_{j+1}^{-1} {\cal L}_{{\cal A} j}(\lambda) M_{j}\label{gaugetransf}
\EN
where $M_{j}$ are arbitrary invertible 2 $\times$ 2 c-number matrices acting on the space ${\cal A}$. After performing this gauge transformations
the transfer matrix (\ref{transferT}) becomes
\EQ
T(\lambda)=Tr_{\cal A}[{ M_{1}^{-1} {\cal G}_{\cal A} M_{L+1} 
\widetilde{{\cal T}}_{{\cal A} } }(\lambda)]
\label{trans}
\EN
where $ \widetilde{{\cal T}}_{{\cal A}}(\lambda)
=\widetilde{{\cal L}}_{{\cal A} L}(\lambda) \dots \widetilde{{\cal L}}_{{\cal A} 1}(\lambda) $.

The next step is to look for gauge transformations $M_{j}$ such that $\widetilde{{\cal L}}_{{\cal A} j}(\lambda)$ is annihilated for
instance by its lower left element for arbitrary values of the spectral parameter. Representing the matrices $M_{j}$ by
\EQ
M_{j}=\left(\begin{array}{cc}
        x_{j} & r_{j} \\
        y_{j} & s_{j} \\
        \end{array}\right)
\EN
we can conclude \cite{BA} that
such annihilation property occurs when the ratio $\frac{x_{j}}{y_{j}}$ is a constant for 
$j=1,\dots,L+1$. As a consequence of that we can choose the local reference state  $\ket{0}_{j}$ as
\EQ
\ket{0}_{j}=\left(\begin{array}{c}
        \frac{x_{j}}{y_{j}} \\
        1 \\
        \end{array}\right)_{j}\label{referencelocal}
\EN
following that the action of the operator $\widetilde{{\cal L}}_{{\cal A}j}(\lambda)$ in this state is given by
\EQ
\widetilde{{\cal L}}_{{\cal A} j}(\lambda) \ket{0}_{j}=\left(\begin{array}{cc}
                        a(\lambda)\frac{y_{j}}{y_{j+1}}\ket{0}_{j} & \# \\
                        0 & b(\lambda)\frac{y_{j+1}}{y_{j}}\frac{det[M_{j}]}{det[M_{j+1}]}\ket{0}_{j} \\
                        \end{array}\right)\label{triangular}
\EN
where the symbol $\#$ represents general non-null values.

The remaining freedom that we have on the matrix elements of $M_{j}$ is now used to choose matrices $M_{1}$ and $M_{L+1}$ in such way
that they transform the boundary matrix ${\cal G}_{{\cal A}}$ 
into a diagonal matrix. More precisely, by imposing that
\EQ
M_{1}^{-1}{\cal G}_{{\cal A}} M_{L+1} =\left(\begin{array}{cc}
                        \widetilde{g}_{1} & 0 \\
                        0 & \widetilde{g}_{2} \\
                        \end{array}\right)\label{conddiag}
\EN
it follows that the constrains for the first column elements are
\bear
g_{11}x_{L+1} + g_{12}y_{L+1} & = &\widetilde{g}_{1}x_{1}
\nonumber \\
g_{21}x_{L+1} + g_{22}y_{L+1} & = &\widetilde{g}_{1}y_{1}
\label{firstcolumn}
\ear
while for the second column elements we have
\bear
g_{11}r_{L+1} + g_{12}s_{L+1} & = &\widetilde{g}_{2}r_{1}
\nonumber \\
g_{21}r_{L+1} + g_{22}s_{L+1} & =  &\widetilde{g}_{2}s_{1}
\label{secondcolumn}
\ear

At this point we emphasize  our assumption that we are dealing with a non-singular boundary matrix. 
While we have an enormous freedom to choose the second column elements the same does not occur for the first ones because the ratio
$\frac{x_{j}}{y_{j}}$ needs to be kept fixed to preserve triangularity of $\widetilde{{\cal L}}_{{\cal A} j}(\lambda)$. This latter fact
together with relation (\ref{firstcolumn}) impose a restriction to this ratio 
which is precisely the same satisfied by ratio of the components
of the eigenvectors of the boundary matrix ${\cal G}_{{\cal A}}$. Therefore, we have two 
possibilities for the ratio $p^{(\pm)}=\frac{x_{j}}{y_{j}}$
which are
\EQ
p^{(\pm)}=\frac{(g_{11}-g_{22}) \pm \sqrt{(g_{11}-g_{22})^2 + 4g_{12}g_{21}}}{2g_{21}}
\label{ppm}
\EN

Putting now all these informations together 
it is possible to build up two appropriate global reference states $\ket{0}^{(\pm)}$ by
the tensor product
\EQ
\ket{0}^{(\pm)}=\prod_{j=1}^{L} \otimes \left(\begin{array}{c}
                                         p^{(\pm)} \\
                                         1 \\
                                         \end{array}\right)_{j}
\label{globalvaccum}
\EN

At this point the state (\ref{globalvaccum}) preserves at least 
the desirable triangular property of the total monodromy $M_{1}^{-1}{\cal G}_{{\cal A}} M_{L+1}
\widetilde{{\cal T}}_{\cal A}(\lambda)$. Below 
we shall see that they are indeed eigenstates of the transfer matrix (\ref{transferT})
independent of further choices of the elements of the gauge matrices $M_{j}$.
Further progress is made by recasting the Yang-Baxter algebra for the gauge transformed monodromy
$\widetilde{{\cal T}}_{{\cal A}}(\lambda)$
in the form of
commutation relations for the creation and annihilation fields. In order to do that it is convenient to represent $\widetilde{\cal T}_{{\cal A}}(\lambda)$
by the following  2 $ \times $ 2 matrix
\EQ
\widetilde{\cal T}_{{\cal A}}(\lambda)=\left(\begin{array}{cc}
                                \widetilde{A}(\lambda) & \widetilde{B}(\lambda) \\
                                \widetilde{C}(\lambda) & \widetilde{D}(\lambda) \\
                                \end{array}\right)
\EN

As a consequence of the triangular property (\ref{triangular}) we are able to derive important relations for the diagonal elements of the
transformed monodromy matrix
\bear
\widetilde{A}(\lambda)\ket{0}^{(\pm)} & = & [a(\lambda)]^{L}\frac{y_{1}}{y_{L+1}}\ket{0}^{(\pm)} \nonumber \\
\widetilde{D}(\lambda)\ket{0}^{(\pm)} & = & [b(\lambda)]^{L}\frac{y_{L+1}}{y_{1}}\frac{det[M_{1}]}{det[M_{L+1}]}\ket{0}^{(\pm)}
\ear
besides the annihilation property
\EQ
\widetilde{C}(\lambda)\ket{0}^{(\pm)}=0
\EN

Now taking into account that gauge matrices $M_{j}$ are themselves 
symmetries allowed by the property (\ref{Gcondition}) it is
not difficult to show that the gauge transformed monodromy 
$\widetilde{{\cal T}}_{{\cal A}}(\lambda)$
matrix satisfies the same Yang-Baxter algebra as the
original monodromy matrix ${\cal T_{\cal{A}}}(\lambda)$. In other words, we have 
that $\widetilde{\cal T}_{{\cal A}}(\lambda)$ satisfies the relation
\EQ
R(\lambda, \mu) \widetilde{{\cal T_{\cal{A}}}}(\lambda) \otimes \widetilde{{\cal T_{\cal{A}}}}(\mu) =
\widetilde{{\cal T_{\cal{A}}}}(\mu) \otimes \widetilde{{\cal T}_{\cal{A}}}(\lambda) R(\lambda, \mu)
\EN
where in our case the $R$-matrix is given by
\EQ
R(\lambda, \mu)= \left(\begin{array}{cccc}
                a(\lambda- \mu) & 0 & 0 & 0 \\
                0 & 1 & b(\lambda- \mu) & 0 \\
                0 & b(\lambda- \mu) & 1 & 0 \\
                0 & 0 & 0 & a(\lambda- \mu) \\
                \end{array}\right)\label{Rmatrix}
\EN

This means that we have the same set of commutation rules 
of the periodic six vertex model \cite{FA,KO} however now for the
gauged matrix elements. Out of sixteen possible relations three of them are of great use, namely
\bear
\widetilde{A}(\lambda)\widetilde{B}(\mu) & = & \frac{a(\mu - \lambda)}{b(\mu -\lambda)} \widetilde{B}(\mu)\widetilde{A}(\lambda)
- \frac{1}{b(\mu - \lambda)}\widetilde{B}(\lambda)\widetilde{A}(\mu) \label{commutN2A} \\
\widetilde{D}(\lambda)\widetilde{B}(\mu) & = & \frac{a(\lambda- \mu)}{b(\lambda- \mu)} \widetilde{B}(\mu)\widetilde{D}(\lambda)
- \frac{1}{b(\lambda- \mu)}\widetilde{B}(\lambda)\widetilde{D}(\mu) \\
\left[\widetilde{B}(\lambda), \widetilde{B}(\mu)\right] & = & 0 \label{commutN2B}
\ear

The fields $\widetilde{B}(\lambda)$ are then interpreted as a 
kind of creation operators  over the pseudovacuum $\ket{0}^{(\pm)}$
and a natural  ansatz for the eigenvectors $\ket{\phi}^{(\pm)}$ of the transfer matrix $T(\lambda)$ is
\EQ
\ket{\phi}^{(\pm)}=\prod_{j=1}^{n_{\pm}}\widetilde{B}(\lambda_{j}^{(\pm)})\ket{0}^{(\pm)}
\label{einv}
\EN

The eigenvalue problem $T(\lambda)\ket{\phi}^{(\pm)}=\Lambda^{(\pm)}(\lambda)\ket{\phi}^{(\pm)}$ now becomes
\EQ
\left[ \widetilde{g}_{1}\widetilde{A}(\lambda) + \widetilde{g}_{2} \widetilde{D}(\lambda)\right] \ket{\phi}^{(\pm)}=
\Lambda^{(\pm)}(\lambda)\ket{\phi}^{(\pm)}
\EN
and it can be solved in the same way as the periodic six vertex model \cite{FA}, i.e 
by taking the fields $\widetilde{A}(\lambda)$ and $\widetilde{D}(\lambda)$
through the creation operators $\widetilde{B}(\lambda)$ with the help of the commutation rules (\ref{commutN2A}-\ref{commutN2B}). One
peculiarity here, however, is that the calculations 
involving the action of the diagonal fields $\widetilde{A}(\lambda)$ and $\widetilde{D}(\lambda)$
over the reference state $\ket{0}^{(\pm)}$ requires extra simplifications to eliminate unnecessary dependence of the gauge matrices elements.
They are carried out by using the help of Eqs.(\ref{firstcolumn} - \ref{ppm}) 
and our final result for the eigenvalues $\Lambda^{(\pm)}(\lambda)$
are
\EQ
\Lambda^{(\pm)}(\lambda)= g^{(\pm)}[a(\lambda)]^{L}\prod_{i=1}^{n_{\pm}}\frac{\lambda_{i}^{(\pm)}-\lambda+\frac{1}{2}}{\lambda_{i}^{(\pm)}-\lambda -\frac{1}{2}}
+ g^{(\mp)}[b(\lambda)]^{L}\prod_{i=1}^{n_{\pm}}\frac{\lambda-\lambda_{i}^{(\pm)}+\frac{3}{2}}{\lambda-\lambda_{i}^{(\pm)}+\frac{1}{2}}
\label{EigenvalueN2}
\EN
provided that the rapidities $\lambda_{i}^{(\pm)}$ satisfy the following Bethe ansatz equations
\EQ
\left[\frac{\lambda_{i}^{(\pm)}+\frac{1}{2}}{\lambda_{i}^{(\pm)}-\frac{1}{2}}\right]^{L}=\frac{g^{(\mp)}}{g^{(\pm)}}
\prod_{\stackrel{j=1}{j \neq i }}^{n_{\pm}} \frac{\lambda_{i}^{(\pm)}-\lambda_{j}^{(\pm)}+1 }{\lambda_{i}^{(\pm)}-\lambda_{j}^{(\pm)}-1}
\label{BAN2}
\EN
where we have performed the convenient shift $\lambda_{i}^{(\pm)} \rightarrow \lambda_{i}^{(\pm)}-\frac{1}{2}$. The phase factors $g^{(\pm)}$ are just the
eigenvalues of the matrix ${\cal G}_{{\cal A}}$
\EQ
g^{(\pm)}=\frac{(g_{11}+g_{22}) \pm \sqrt{(g_{11}-g_{22})^2 + 4g_{12}g_{21}}}{2}
\EN

Rather remarkably, we see that the 
final form of the eigenvalues and Bethe ansatz equations resemble much that of the isotropic six vertex
model with diagonal boundary if we replace the 
diagonal twists by the eigenvalues of the non-diagonal boundary ${\cal G}_{\cal A}$ matrix.
The eigenvectors can also be thought as multi-particle states in which the integers $n_{\pm} \leq L$ play 
the role of particle number sectors.
We emphasize, however, that the corresponding basic creation fields are much more sophisticated operators than that of the periodic six vertex
model \cite{FA}. It is tempting to think that the two possible ways we have at our disposal to build up the Hilbert space is related to the
remaining $Z_{2}$ symmetry allowed by boundary terms. One expects therefore that it should be possible to obtain the eigenvalues of the
transfer matrix either from the $\ket{0}^{(+)}$ or $\ket{0}^{(-)}$ pseudovacuums. Indeed, we have verified this fact by numerically solving the
equations for some values of $L$ and comparing them to exact diagonalization of the transfer matrix (\ref{transferT}). We note, however, that
a given eigenvalue of the transfer matrix is in general obtained 
at different particle sectors $n_{\pm}$ over the $\ket{0}^{(\pm)}$ reference
states. For example, the eigenvalue $g^{+} [a(\lambda)]^{L}+g^{-}[b(\lambda)]^{L}$  can be obtained 
either from the  zero-particle state
$\ket{0}^{(+)}$ or as 
a $L$-particle state over the pseudovacuum $\ket{0}^{(-)}$. 
In Appendix A, we present details of our study for $L=2$ in which 
Eqs.(\ref{EigenvalueN2}-\ref{BAN2}) can be solved by analytical means. 
Our numerical results up to $L=4$ suggest that two possible branches of the  Bethe 
ansatz solutions (\ref{BAN2}) produce
the complete spectrum of the transfer matrix (\ref{transferT}). It would be interesting to 
further investigate the completeness of the Bethe ansatz
(\ref{BAN2}) by adapting the recent arguments developed by Baxter \cite{BA1} to the case of non-diagonal twists.

We now can derive similar results for the spin chain 
that commutes with the transfer matrix (\ref{transferT}). The corresponding spin-$1/2$
XXX Hamiltonian follows from expression (\ref{HAMIL})  and it is given by
\EQ
{\cal H}={\cal J}\sum_{j=1}^{L}\left( \sigma_{j}^{+}\sigma_{j+1}^{-}+ \sigma_{j}^{-}\sigma_{j+1}^{+} +\frac{\sigma_{j}^{z}\sigma_{j+1}^{z}}{2} \right)
\EN
with the following boundary condition
\EQ
\left(\begin{array}{c}
        \sigma_{L+1}^{+} \\
        \sigma_{L+1}^{-} \\
        \sigma_{L+1}^{z}
        \end{array}\right)=
        \frac{1}{g_{11}g_{22}-g_{12}g_{21}}\left( \begin{array}{ccc}
        g_{11}^{2} & -g_{21}^{2} & - g_{11}g_{21} \\
        -g_{12}^{2} & g_{22}^{2} & g_{12}g_{22} \\
        -2g_{11}g_{12} & 2g_{21}g_{22} & g_{11}g_{22}+g_{12}g_{21}
                \end{array} \right)\left(\begin{array}{c}
                                \sigma_{1}^{+} \\
                                \sigma_{1}^{-} \\
                                \sigma_{1}^{z}
                                \end{array}\right)
\EN

Its eigenvalues 
$E^{(\pm)}=\frac{d Log[\Lambda^{(\pm)}(\lambda)]}{d\lambda}\mid_{\lambda=0}$  are
\EQ
E^{(\pm)}={\cal J}\sum_{j=1}^{n_{\pm}} \frac{1}{\lambda_{j}^{\pm 2}-\frac{1}{4}}+\frac{{\cal J}L}{2}
\EN
where $\lambda_i^{(\pm)}$ satisfy the Bethe ansatz equations  (\ref{BAN2}). 

Our final comment concerns with the comparison between our results (\ref{EigenvalueN2}-\ref{BAN2}) and that
of refs.\cite{BAT,BAT1} in the isotropic limit case when the trigonometric weights become rational
functions. We see that they are in accordance for the common non-diagonal boundary  
$ {\cal G}_{\cal A}=\left( \begin{array}{cc}
        0 &   g_{12} \\
        g_{21} &   0 \\
        \end{array}\right) $ apart from the fact that our numbers of roots $n_{\pm}$ can vary up
to $L$ while that of refs.\cite{BAT,BAT1} are fixed at $L$. This implies that for non-diagonal
boundary conditions the complete solution of the isotropic limit 
does not follows directly from that found for the anisotropic six vertex model \cite{BAT,BAT1}.
This means that even for this particular non-diagonal boundary  
the results (\ref{EigenvalueN2}-\ref{BAN2}) are novel in the literature.

\section{Nested Bethe ansatz for $SU(N)$ model}\label{Nested Bethe ansatz for SUN model}

The purpose of this section is to generalize the results of the previous section for general
$N$. We wish to diagonalize the transfer matrix (\ref{transferT}), where now the operator
${\cal L}_{{\cal A}i}(\lambda)$ is
\EQ
{\cal L}_{{\cal A} i}(\lambda)=a(\lambda)\sum_{\alpha=1}^{N} \hat{e}_{\alpha\alpha}^{(\cal A)} \otimes \hat{e}_{\alpha\alpha}^{(i)}
+ b(\lambda)\sum_{\stackrel{\alpha,\beta=1}{\alpha \neq \beta}}^{N} \hat{e}_{\alpha\alpha}^{(\cal A)} \otimes \hat{e}_{\beta\beta}^{(i)}
+ \sum_{\stackrel{\alpha,\beta=1}{\alpha \neq \beta}}^{N} \hat{e}_{\alpha\beta}^{(\cal A)} \otimes \hat{e}_{\beta\alpha}^{(i)} \label{laxN}
\EN
where $\hat{e}_{ij}^{(V)}$ are the standard Weyl matrices whose
elements acting on the space $V$ are $[\hat{e}_{ij}^{(V)}]_{kl}=\delta_{ik}\delta_{jl}$. In this basis
the boundary matrix ${\cal G}_{{\cal A}}$ is generally represented by
\EQ
{\cal G}_{\cal A}=\sum_{\alpha,\beta=1}^{N} g_{\alpha\beta}\hat{e}_{\alpha\beta}^{({\cal A})}
\EN

As before we have to seek for suitable references states by imposing the gauge transformation
(\ref{gaugetransf}) for each operator (\ref{laxN}) and require that they are
up triangular when acting on such pseudovacuum. Denoting the gauge matrices by
$ M_{j}= \displaystyle \sum_{\alpha,\beta=1}^{N} m_{j}(\alpha,\beta) \hat{e}_{\alpha\beta}^{(\cal A)}$ we find that such
triangular property is fully achieved when the following ratios relations are satisfied
\EQ
p_{\alpha,\beta}=\frac{m_{j}(\alpha,\beta)}{m_{j}(N,\beta)}=\frac{m_{j+1}(\alpha,\beta)}{m_{j+1}(N,\beta)},
 ~~ \alpha,\beta=1,\dots, N-1
\label{constrain}
\EN
for each $j=1,\dots,L+1$. In terms of these ratios the local reference state $\ket{0}_{j}$ assume the form
\EQ
\ket{0}_{j}=\left(\begin{array}{c}
                p_{1,1} \\
                p_{2,1} \\
                \vdots \\
                p_{N-1,1} \\
                1
                \end{array}\right)_j
\label{VEC}
\EN

Other important ingredient is the
action of the gauge transformed operator $\widetilde{\cal L}_{{\cal A}j}(\lambda)$ over the local
state of reference. This now can be  represented by the following
$N \times N$ matrix on the space ${\cal A}$
\EQ
\widetilde{{\cal L}}_{{\cal A} j} \ket{0}_{j}=\left(\begin{array}{ccccc}
                        a(\lambda)\frac{f_{1}^{j}}{f_{1}^{j+1}} \ket{0}_{j} & \# & \# & \cdots & \# \\
                        0 & b(\lambda)\frac{f_{2}^{j}}{f_{2}^{j+1}}\ket{0}_{j} & \# & \cdots & \# \\
                        \vdots & \vdots & \vdots & \ddots & \vdots \\
                        0 & 0 & 0 & \cdots & b(\lambda) \frac{f_{N}^{j}}{f_{N}^{j+1}}\ket{0}_{j}
                        \end{array}\right)_{N \times N}
\label{laxNtriang}
\EN
where the variables $f_{\alpha}^{j}$ are given by
\EQ
f_{\alpha}^{j}= \cases{
\displaystyle  m_{j}(N,\alpha), ~~ \alpha=1,\dots ,N-1 \cr
\displaystyle \left(\prod_{i=1}^{N-1} \frac{1}{m_{j}(N,i)}\right) det[M_{j}], ~~ \alpha=N \cr }
\EN

Similarly to the previous section we can take advantage of the 
remaining freedom of the elements of the gauge matrices to transform
$M_{1}^{-1}{\cal G}_{{\cal A}} M_{L+1}$ into a diagonal matrix. By imposing this condition
the matrices elements of $M_1$ and $M_{L+1}$ become related by the expression
\EQ
\sum_{\gamma=1}^{N} g_{\alpha\gamma} m_{L+1}(\gamma,\beta)=m_{1}(\alpha,\beta)\widetilde{g}_{\beta}, ~~
\alpha,\beta=1,\dots,N\label{constr}
\EN
where $\widetilde{g}_{\alpha}$ represent the  diagonal elements of the transformed boundary matrix.

Equations (\ref{constrain}) and (\ref{constr}) together impose constrains to the possible values
ratios $p_{\alpha,\beta}$  which turns out to be same conditions satisfied by the 
ratio of the components of the eigenvectors of boundary matrix ${\cal G}_{{\cal A}}$.  
This means that we have $N$ possible choices for $p_{\alpha,1}^{(l)}$~$l=1,\dots,N$ and
consequently from
Eq.(\ref{VEC}) $N$ kind of suitable local references states  
$\ket{0}_{j}^{(l)}$. A natural ansatz for the $N$ possible choices of global reference states 
are
\EQ
\ket{0}^{(l)}=\prod_{j=1}^{L} \otimes 
\ket{0}_j^{(l)} 
~~l=1,\dots,N 
\EN

The next step is to write a suitable representation for the gauge transformed monodromy
matrix in the auxiliary space ${\cal A}$. The triangular property (\ref{laxNtriang}) suggests
us to seek for the structure used in nested Bethe ansatz diagonalization of the periodic
$SU(N)$ vertex models \cite{KUL,DE} which is
\EQ
\widetilde{{\cal T}}(\lambda)=\left(\begin{array}{cccc}
                \widetilde{A}(\lambda) & \widetilde{B}_{1}(\lambda) & \cdots & \widetilde{B}_{N-1}(\lambda) \\
                \widetilde{C}_{1}(\lambda) & \widetilde{D}_{11}(\lambda) & \cdots & \widetilde{D}_{1N-1}(\lambda) \\
                \vdots & \vdots & \ddots & \vdots \\
                \widetilde{C}_{N-1}(\lambda) & \widetilde{D}_{N-11}(\lambda) & \cdots & \widetilde{D}_{N-1 N-1}(\lambda) \\
                \end{array}\right)_{N \times N}
\EN

The triangularity property (\ref{laxNtriang}) implies that the fields
$\widetilde{B}_{i}(\lambda)$ play the role of creations operators, $\widetilde{C}_{i}(\lambda)$
are annihilation fields while the diagonal operator $\widetilde{A}(\lambda)$ and
$\widetilde{D}_{ii}(\lambda)$ acts on the reference state $\ket{0}^{(l)}$ as
\bear
\widetilde{A}(\lambda)\ket{0}^{(l)} & = & [a(\lambda)]^{L} \frac{f_{1}^{1}}{f_{1}^{L+1}} \ket{0}^{(l)} \\
\widetilde{D}_{ii}(\lambda)\ket{0}^{(l)} & =& [b(\lambda)]^{L}\frac{f_{i+1}^{1}}{f_{i+1}^{L+1}}\ket{0}^{(l)},~i=1,
\dots,N-1
\ear

To construct other eigenvectors we shall use the commutation relations between the gauge
transformed monodromy matrix elements.  The arguments used in section 2 allows to
conclude that these 
commutation rules are the same as that already known for the periodic
$SU(N)$ models \cite{KUL,DE}. 
The most useful relations for subsequent derivations are
\bear
\widetilde{A}(\lambda)\widetilde{B}_{i}(\mu) = \frac{a(\mu-\lambda)}{b(\mu-\lambda)}\widetilde{B}_{i}(\mu)\widetilde{A}(\lambda)
-\frac{1}{b(\mu-\lambda)}\widetilde{B}_{i}(\lambda)\widetilde{A}(\mu) \label{COMM} \\
\widetilde{D}_{ij}(\lambda)\widetilde{B}_{k}(\mu) 
=\frac{1}{b(\lambda-\mu)}\widetilde{B}_{p}(\mu)\widetilde{D}_{iq}(\lambda)r^{(1)}(\lambda-\mu)_{pq}^{jk}
-\frac{1}{b(\lambda-\mu)}\widetilde{B}_{j}(\lambda)\widetilde{D}_{ik}(\mu) \label{commutD} 
\label{COMM1}
\ear
\EQ
\widetilde{B}_{i}(\lambda)\widetilde{B}_{j}(\mu) = \widetilde{B}_{p}(\mu)\widetilde{B}_{q}(\lambda)r^{(1)}(\lambda-\mu)_{pq}^{ij}
\label{COMM2}
\EN
where $r^{(1)}(\lambda)_{pq}^{ij}$ are the elements of the $R$-matrix  associated to the $SU(N-1)$ vertex model.

In terms of the gauge transformed fields, the eigenvalue problem for the transfer matrix $T(\lambda)$ becomes
\EQ
\left[\widetilde{g}_{1}\widetilde{A}(\lambda) + 
\sum_{i=1}^{N-1}\widetilde{g}_{i+1} \widetilde{D}_{ii}(\lambda)\right]\ket{\phi}^{(l)}=
\Lambda^{(l)}(\lambda)\ket{\phi}^{(l)}
\label{TtransN}
\EN
where $\ket{\phi}^{(l)}$ denotes the eigenvectors. Previous experience with these models \cite{KUL,DE} 
suggests us to suppose that eigenvectors can be written in terms of the following linear combination
\EQ
\ket{\phi}^{(l)}=\widetilde{B}_{a_{1}}(\lambda_{1}^{(1,l)})\dots 
\widetilde{B}_{a_{m_{1}^{l}}}(\lambda_{m_{1}^{l}}^{(1,l)})
{\cal F}^{a_{m_{1}^{l}} \dots a_{1}} \ket{0}^{(l)}
\label{MULTI}
\EN
where sum over repeated indices $a_{n}=1, \dots N-1 $ is assumed. At this stage the components 
${\cal F}^{a_{m_{1}^{l}}\dots a_{1}}$ are thought as coefficients of an arbitrary linear combination
that are going to be determined a posteriori.

By carrying on the fields $\widetilde{A}(\lambda)$ and $\widetilde{D}_{ii}(\lambda)$ over the
multi-particle state (\ref{MULTI}) we generate terms that are 
proportional to $\ket{\phi}^{(l)}$ and those that are not the so-called unwanted terms. The first ones will
contribute directly to the eigenvalue $\Lambda^{(l)}(\lambda)$ and are obtained by keeping only the first term
of the commutation rules  
(\ref{COMM}-\ref{COMM1}). These calculations are by now standard in the literature and here we present
only the main results of the action of the transfer matrix on the eigenvector $\ket{\phi}^{(l)}$ which is 
\bear
T(\lambda)
\ket{\phi}^{(l)} & = &
\widetilde{g}_{1}\frac{f_{1}^{1}}{f_{1}^{L+1}} [a(\lambda)]^{L} \prod_{j=1}^{m_{1}^{l}}
\frac{a(\lambda_{j}^{(1,l)}-\lambda)}{b(\lambda_{j}^{(1,l)}-\lambda)}\ket{\phi}^{(l)}
\nonumber \\
&& +[b(\lambda)]^{L}
\prod_{j=1}^{m_{1}^{l}}\frac{1}{b(\lambda-\lambda_{j}^{(1,l)})}
\widetilde{B}_{b_{1}}(\lambda_{1}^{(1,l)}) \dots \widetilde{B}_{b_{m_{1}^{l}}}(\lambda_{m_{1}^{l}}^{(1,l)})
T^{(1)}(\lambda, \{ \lambda_{i}^{(1,l)} \})_{b_{1} \dots b_{m_{1}^{l}}}^{a_{1} \dots a_{m_{1}^{l}}} 
{\cal F}^{a_{m_{1}^{l}}
\dots a_{1}} \ket{0}^{(l)}
\nonumber \\
&& + \mathrm{ unwanted~ terms}
\label{eigeneqN}
\ear

All the pieces entering the above expression can be summarized as follows. The terms 
$ T^{(1)}(\lambda, \{\lambda_{i}^{(1,l)}\})_{b_{1} \dots b_{m_{1}^{l}}}^{a_{1} \dots a_{m_{1}^{l}}} $ are 
transfer matrix elements of an auxiliary inhomogeneous problem related to the $SU(N-1)$ vertex model with
twisted boundaries $\widetilde{\cal{G}}$ defined by
\EQ
T^{(1)}(\lambda, \{\lambda_{i}^{(1,l)}\})_{b_{1} \dots b_{m_{1}^{l}}}^{ a_{1} \dots a_{m_{1}^{l}}} =
r^{(1)}(\lambda-\lambda_{1}^{(1,l)})_{b_{1} d_{1}}^{a a_{1}} 
r^{(1)}(\lambda-\lambda_{2}^{(1,l)})_{b_{2}d_{2}}^{d_{1} a_{2}} 
\dots r^{(1)}(\lambda-\lambda_{m_{1}^{l}}^{(1,l)})_{b_{m_{1}^{l}} d_{m_{1}^{l}}}^{d_{m_{1}^{l}-1} a_{m_{1}^{l}}}
\widetilde{\cal{G}}_{a d_{m_1^l}}
\EN
where 
$\widetilde{\cal{G}}_{ab}$ denotes the elements of the boundary matrix $\widetilde{\cal{G}}$ given by 
\EQ
\widetilde{\cal{G}}=\left(\begin{array}{ccccc}
                        \widetilde{g}_{2}\frac{f_{2}^{1}}{f_{2}^{L+1}} & \# & \# & \cdots & \# \\
                        0 & \widetilde{g}_{3} \frac{f_{3}^{1}}{f_{3}^{L+1}} & \# & \cdots & \# \\
                        \vdots & \vdots & \vdots & \ddots & \vdots \\
                        0 & 0 & 0 & \cdots & \widetilde{g}_{N} \frac{f_{N}^{1}}{f_{N}^{L+1}}
                        \end{array}\right)_{N-1 \times N-1}
\label{bound}
\EN

The unwanted terms are originated when the variables  
$\lambda^{(1,l)}_i$ of the multi-particle state (\ref{MULTI}) are exchanged with the spectral 
parameter $\lambda$ due to the second part of the commutation rules (\ref{COMM}-\ref{COMM1}). It is possible 
to collect these terms in closed forms, thanks to the commutation rule (\ref{COMM2})
which makes possible to relate different ordered multi-particle states. It turns out that all the unwanted terms
are canceled out provided that the rapidities 
$\lambda_{i}^{(1,l)}$ satisfy the following restriction,
\bear
&& \widetilde{g}_{1} \frac{f_1^1}{f_1^{L+1}} \left[\frac{a(\lambda^{(1,l)}_{i})}  
{b(\lambda^{(1,l)}_{i})} \right]^{L}
\prod_{\stackrel{j=1}{j \neq i}}^{m_{1}^{l}} b(\lambda^{(1,l)}_{i}-\lambda^{(1,l)}_{j})
\frac{a(\lambda^{(1,l)}_{j}-\lambda^{(1,l)}_{i})}{b(\lambda^{(1,l)}_{j}-\lambda^{(1,l)}_{i})}
{\cal F}^{a_{m_{1}^l} \dots a_{1}}=
\nonumber \\
&& T^{(1)}(\lambda=\lambda^{(1,l)}_{i},\{\lambda^{(1,l)}_{j}\})^{b_{1} \cdots b_{m_{1}^l}}_{a_{1} \cdots a_{m_{1}^l}} 
{\cal F}^{b_{m_{1}^l} \dots b_{1}} , i=1, \dots, m_{1}^l
\label{BABA}
\ear

Now we reached a point which is fundamental to diagonalize  
$T^{(1)}(\lambda, \{\lambda_{i}^{(1,l)}\})$ in order to compute the eigenvalues of $T(\lambda)$ and at the same
time to solve Eq.(\ref{BABA}). This becomes possible if we require that 
${\cal F}^{a_{m_{1}^l} \dots a_{1}}$ is an eigenvector of  the auxiliary  transfer matrix with eigenvalue 
$\Lambda^{(1)}(\lambda,\{\lambda^{(1,l)}_{i}\})$, namely
\EQ
T^{(1)}(\lambda,\{\lambda^{(1,l)}_{i}\})^{b_{1} \cdots b_{m_{1}^l}}_{a_{1} \cdots a_{m_{1}^l}} {\cal F}^{b_{m_{1}^l} \dots b_{1}} =
\Lambda^{(1)}(\lambda,\{\lambda^{(1)}_{i}\}) {\cal F}^{a_{m_{1}^l} \dots a_{1}}
\label{NEST}
\EN

Inspection of Eq.(\ref{eigeneqN}) and Eq.(\ref{BABA}) together with Eq.(\ref{NEST}) shows that the
eigenvalue of $T(\lambda)$ is
\EQ
\Lambda(\lambda)^{(l)}  =
\widetilde{g}_{1} 
\frac{f_1^1}{f_1^{L+1}} 
[a(\lambda)]^{L} 
\prod_{i=1}^{m_{1}^l} \frac{a(\lambda^{(1,l)}_{i}-\lambda)}
{b(\lambda^{(1,l)}_{i}-\lambda)} 
+[b(\lambda)]^{L} \prod_{i=1}^{m_{1}^l}
\frac{1}{b(\lambda-\lambda^{(1,l)}_{i})}
\Lambda^{(1)}(\lambda,\{\lambda^{(1,l)}_{i}\})  
\EN
and the nested Bethe ansatz equations (\ref{BABA}) become
\EQ
\widetilde{g}_{1} 
\frac{f_1^1}{f_1^{L+1}} 
\left[ \frac{a(\lambda^{(1,l)}_{i})}  
{b(\lambda^{(1,l)}_{i})} \right]^{L}
\prod_{\stackrel{j=1}{j \neq i}}^{m_{1}^l} b(\lambda^{(1,l)}_{i}-\lambda^{(1,l)}_{j})
\frac{a(\lambda^{(1,l)}_{j}-\lambda^{(1,l)}_{i})}{b(\lambda^{(1,l)}_{j}-\lambda^{(1,l)}_{i})}  = 
\Lambda^{(1)}(\lambda = \lambda^{(1,l)}_{i},\{\lambda^{(1,l)}_{j}\})
\nonumber \\
,~~ i=1, \dots, m_{1}^l
\EN

In order to solve the eigenvalue problem (\ref{NEST}) it is necessary to introduce a second 
algebraic Bethe
ansatz for the eigenvectors 
${\cal F}^{a_{m_{1}^l} \dots a_{1}}$. Because the boundary matrix $\widetilde{\cal{G}}$ is triangular 
there is no need to perform gauge transformations to find an appropriate reference state  for
$ T^{(1)}(\lambda,\{\lambda^{(1,l)}_{i}\})$.  We can use, for instance, the usual ferromagnetic
pseudovacuum build up by tensor product of  elementary $(N-1)$-dimensional 
$\pmatrix{
1 \cr
0 \cr
\vdots \cr
0 \cr}_{N-1} $ vectors. As a result the solution (\ref{NEST}) becomes very similar to that of the periodic
$SU(N-1)$ vertex model in the presence of inhomogeneities. Since this problem has been
extensively discussed in the literature we will only present our final results for the main eigenvalue problem 
(\ref{TtransN}). It turns out that the eigenvalues of the transfer matrix $\Lambda^{(l)}(\lambda)$ is given by
\bear
\Lambda^{(l)}(\lambda;\{\lambda_{i}^{(1,l)}\}, \dots, \{\lambda_{i}^{(N-1, l)}\})&  =&
\widetilde{g}_{1} \frac{f_{1}^{1}}{f_{1}^{L+1}} [a(\lambda)]^{L} \prod_{j=1}^{m_{1}^{l}} \frac{a(\lambda_{j}^{(1, l)}-\lambda)}{b(\lambda_{j}^{(1, l)}-\lambda)} \nonumber \\
&& +[b(\lambda)]^{L} \sum_{k=1}^{N-2} \widetilde{g}_{k+1} \frac{f_{k+1}^{1}}{f_{k+1}^{L+1}} 
\prod_{j=1}^{m_{k}^{l}}\frac{a(\lambda-\lambda_{j}^{(k, l)})}{b(\lambda-\lambda_{j}^{(k, l)})}
\prod_{j=1}^{m_{k+1}^{l}} \frac{a(\lambda_{j}^{(k+1, l)}-\lambda)}{b(\lambda_{j}^{(k+1, l)}-\lambda)}
\nonumber \\
&& +[b(\lambda)]^{L} \widetilde{g}_{N} \frac{f_{N}^{1}}{f_{N}^{L+1}} 
\prod_{j=1}^{ m_{N-1}^{l} }\frac{a(\lambda-\lambda_{j}^{(N-1, l)})}{b(\lambda-\lambda_{j}^{(N-1, l)})}
\label{presque}
\ear

The rapidities $\{ \lambda_i^{(k,l)} \}$ $k=1,\dots,N$ parameterize the multi-particle states of the nesting
problem at step $k$ and are required to satisfy the following nested Bethe ansatz equations
\EQ
\frac{\widetilde{g}_{1}}{\widetilde{g}_{2}} \frac{f_{1}^{1} f_{2}^{L+1}}{f_{1}^{L+1} f_{2}^{1}} \left[\frac{a(\lambda_{i}^{(1, l)})}{b(\lambda_{i}^{(1, l)})}\right]^{L}=
\prod_{\stackrel{j=1}{j \neq i}}^{m_{1}^{l}} -\frac{a(\lambda_{i}^{(1, l)}-\lambda_{j}^{(1, l)})}{a(\lambda_{j}^{(1, l)}-\lambda_{i}^{(1, l)})}
\prod_{\stackrel{j=1}{j \neq i}}^{m_{2}^{l}} \frac{a(\lambda_{j}^{(2, l)}-\lambda_{i}^{(1, l)})}{b(\lambda_{j}^{(2, l)}-\lambda_{i}^{(1, l)})}
\EN
\bear
\frac{\widetilde{g}_{k}}{\widetilde{g}_{k+1}} \frac{f_{k}^{1} f_{k+1}^{L+1}}{f_{k}^{L+1} f_{k+1}^{1} } \prod_{j=1}^{m_{k-1}^{l}}\frac{ a(\lambda_{i}^{(k, l)}-\lambda_{j}^{(k-1, l)})}{ b(\lambda_{i}^{(k, l)}-\lambda_{j}^{(k-1, l)})}=
\prod_{\stackrel{j=1}{j \neq i}}^{m_{k}^{l}} -\frac{a(\lambda_{i}^{(k, l)}-\lambda_{j}^{(k, l)})}{a(\lambda_{j}^{(k, l)}-\lambda_{i}^{(k, l)})}
\prod_{\stackrel{j=1}{j \neq i}}^{m_{k+1}^{l}}\frac{a(\lambda_{j}^{(k+1, l)}-\lambda_{i}^{(k, l)})}{b(\lambda_{j}^{(k+1, l)}-\lambda_{i}^{(k, l)})}, \\
 k=2, \dots, N-2 \nonumber
\ear
\EQ
\frac{\widetilde{g}_{N-1}}{\widetilde{g}_{N}} \frac{f_{N-1}^{1} f_{N}^{L+1}}{f_{N-1}^{L+1} f_{N}^{1} } 
\prod_{\stackrel{j=1}{j \neq i}}^{m_{N-2}^{l}} \frac{a(\lambda_{i}^{(N-1, l)}-\lambda_{j}^{(N-2, l)})}{b(\lambda_{i}^{(N-1, l)}-\lambda_{j}^{(N-2, l)})} =
\prod_{\stackrel{j=1}{j \neq i}}^{m_{N-1}^{l}} -\frac{a(\lambda_{i}^{(N-1, l)}-\lambda_{j}^{(N-1, l)})}{a(\lambda_{j}^{(N-1, l)}-\lambda_{i}^{(N-1, l)})}
\label{lasteq}
\EN

The final step is to carry out simplifications on the phase factors $\widetilde{g}_i \frac{f^{1}_i}{f^{L+1}_i}$ with
the help of the constraints (\ref{constrain}) and (\ref{constr}). After a cumbersome algebra it is possible to show that
such factors are just the eigenvalues $g^{(i)}$ of the boundary matrix $\cal{G}_{\cal{A}}$. To make sure that the
different possibilities we have at hand for the ratios $p_{\alpha,\beta}^{(l)}$ do not lead us to singular
gauge matrices we choose to order them for each $l$-th choice of 
pseudovacuum by $g^{(l)}=g^{(l+N)}$ for $l=1,\dots,N$. Taking into
account this ordering as well as performing the shifts $\{ \lambda_j^{(k,l)} \} \rightarrow 
\{ \lambda_j^{(k,l)} \} -k/2$  our result (\ref{presque}) for the eigenvalue becomes
\bear
\Lambda^{(l)}(\lambda;\{\lambda_{i}^{(1, l)}\}, \dots, \{\lambda_{i}^{(N-1, l)}\}) & = &
g^{(l)} [a(\lambda)]^{L} \prod_{j=1}^{m_{1}^{l}}
\frac{\lambda_{j}^{(1, l)}-\lambda+\frac{1}{2}}{\lambda_{j}^{(1, l)}-\lambda -\frac{1}{2}} \nonumber \\
&& +[b(\lambda)]^{L} \sum_{k=1}^{N-2} g^{(l+k)}  \prod_{j=1}^{m_{k}^{l}}
\frac{\lambda-\lambda_{j}^{(k, l)}+\frac{k+2}{2}}{\lambda-\lambda_{j}^{(k, l)}+\frac{k}{2}}
\prod_{j=1}^{m_{k+1}^{l}}
\frac{\lambda_{j}^{(k+1, l)}-\lambda+\frac{1-k}{2}}{\lambda_{j}^{(k+1, l)}-\lambda -\frac{k+1}{2}}
\nonumber \\
&& +[b(\lambda)]^{L} {g}^{(l+N-1)}  \prod_{j=1}^{ m_{N-1}^{l} }
\frac{\lambda-\lambda_{j}^{(N-1,l)}+\frac{N+1}{2}}{\lambda-\lambda_{j}^{(N-1,l)}+\frac{N-1}{2}}
\ear
and the nested Bethe ansatz equations can be compactly written as
\EQ
\left[
\frac{\lambda^{(a,l)}_{i} +\frac{\delta_{a,1}}{2}}{\lambda^{(a,l)}_{i} -\frac{\delta_{a,1}}{2}} 
\right]^{L} = \frac{g^{(l+a)}}{g^{(l+a-1)}}
\prod_{b=1}^{N-1} \prod_{\stackrel{k=1}{k \neq i}}^{m_{b}^l}
\frac{\lambda^{(a,l)}_{i}-\lambda^{(b,l)}_{k} +\frac{C_{a,b}}{2}}{\lambda^{(a,l)}_{i}-\lambda^{(b,l)}_{k} -\frac{C_{a,b}}{2}}, ~~ i=1, \dots, m_{a}^l ;~~ a=1, \dots,N-1 
\EN
where $C_{ab}$ is the Cartan matrix elements of the $SU(N)$ Lie algebra.

We see that the results for the eigenvalues and the Bethe ansatz equations is similar to that expected
from the $SU(N)$ vertex model with diagonal twists giving by the eigenvalues of the boundary matrix $\cal{G}_{\cal{A}}$.
It remains to be investigated whether this interesting feature is particular of the $SU(N)$ symmetry or it also works 
in other 
isotropic vertex models such as those invariant by the 
$O(N)$ and $Sp(2N)$ Lie algebras.

\section{Concluding remarks}\label{Conclusion}

In this paper we have been able to apply the quantum inverse scattering program to solve exactly the isotropic $SU(N)$ vertex model with
non-diagonal twisted boundary conditions. We find that the eigenvectors can be constructed in terms of multi-particle states over $N$ possible
pseudovacuums. The Bethe ansatz results for the eigenvalues are similar to that of the $SU(N)$ model with diagonal boundaries in which the
eigenvalues of the boundary matrix ${\cal G}_{{\cal A}}$ play the role of the diagonal twists.

We expect that our results can be generalized without further difficulties to accommodate the solution of vertex models based on the
$SL(N|M)$ super Lie algebra \cite{SU,PS,KU2} with general non-diagonal twists. These will include interesting systems of correlated
electrons on a lattice such as the one-dimensional supersymmetric t-J model \cite{tJ} and the so-called  Essler, Korepin and Schoutens
superconducting model \cite{EKS} with arbitrary symmetry breaking boundary conditions. With more effort we hope that our approach can be
further generalized to include the trigonometric deformation of those vertex models based on the $U_{q}[SL(N|M)]$ symmetry. In these cases,
however, we recall that the possible ${\cal G}_{\cal A}$ matrices compatible with integrability belong to a smaller group formed by one-dimensional
dilatations and the discrete $Z_{N+M}$ symmetry. 

Other interesting issue that deserves investigation is the situation when the boundary matrix $\cal{G}_{\cal{A}}$ 
becomes singular. For example, one would like to ask it is still possible to exhibit
eigenvectors of the transfer matrix that are given by direct tensor product of $N$-dimensional vectors such
as the reference states of sections 2 and 3. We have studied this problem in the simplest case $N=2$ and
surprisingly we found a family of such states $\ket{\phi}^{(n)}$ which are 
\EQ
\ket{\phi}^{(n)} = 
\prod_{i=1}^{n} \otimes
\pmatrix{
-\frac{g_{22}}{g_{21}} \cr
1 \cr}_i
\prod_{i=n+1}^{L} \otimes
\pmatrix{
\frac{g_{11}}{g_{21}} \cr
1 \cr}_i
,~~n=0,1,\dots,L
\label{STA}
\EN
whose corresponding eigenvalues $\Lambda^{(n)}(\lambda)$ have also the following simple factorized form
\EQ
\Lambda^{(n)}(\lambda)= (g_{11}+g_{22}) [a(\lambda)]^{L-n} [b(\lambda)]^n
\label{JYT}
\EN

This result prompted us to study further properties of the transfer matrix (7) 
when $\cal{G}_{\cal{A}}$ is a singular matrix.
Our study for finite $L$ up to six sites reveals that the roots of the characteristic polynomial of $T(\lambda)$
are exactly the eigenvalues (\ref{JYT})  whose degeneracy is the binomial coefficient
$d_n= \frac{L!}{(L-n)! n!}$. In the case of singular boundary matrix $T(\lambda)$ becomes  defective
since it has fewer than $2^L$
independent eigenvectors. To each distinct eigenvalue $\Lambda^{(n)}(\lambda)$ we find only
one eigenvector which is precisely the state (\ref{STA}) and therefore the total number of
independent states is $L+1$. These results are strong evidences
that  $T(\lambda)$ behaves as a non-derogatory matrix and we conjecture that its Jordan
decomposition for arbitrary $L$ should be 
\EQ
T(\lambda)= diag(J_0,J_1,\cdots,J_L)
\EN
where $J_n$ is a $d_n \times d_n$ Jordan matrix is given by
\EQ
J_n=\left(\begin{array}{ccccc}
                        \Lambda^{(n)}(\lambda) & 1 & 0 & \cdots & 0 \\
                        0 & \Lambda^{(n)}(\lambda) & 1 & \cdots & 0 \\
                        \vdots & \vdots & \vdots & \ddots & \vdots \\
                        0 & 0 & 0 & \cdots & \Lambda^{(n)}(\lambda)
                        \end{array}\right)_{d_n \times d_n}
\label{bound}
\EN

This turns out to be a remarkable example how boundary conditions can change in a drastic way the
Hilbert space of integrable models. At this point it is natural to ask  
what happens to the 
Bethe ansatz states (\ref{einv}) when one gradually varies the boundary matrix towards the singular manifold.
In particular, if we can figure out the kind of Bethe states in each sector $n_{\pm}$ that 
should collapse to the eigenvectors (\ref{STA}). A precise answer 
to this question as well as possible generalizations
of these results for arbitrary $N$  has eluded us so far.

\section*{Acknowledgements}
We would like to thank  the organizers of 
the Workshop Flux, Charge, Topology and Statistics 2003, Amsterdam,
where part of this work was carried out.
The authors G.A.P Ribeiro and W. Galleas thank  Fapesp (Funda\c c\~ao de Amparo \`a Pesquisa
do Estado de S\~ao Paulo) for financial support. The work of M.J. Martins has been partially
support by the Brazilian Research Council-CNPq and Fapesp.

\section*{\bf Appendix A : Completeness for $L=2$}
\setcounter{equation}{0}
\renewcommand{\theequation}{A.\arabic{equation}}

This appendix is concerned with the study of the 
completeness of the Bethe ansatz solution (\ref{BAN2}) for $L=2$, i.e. that all
four eigenvalues of the transfer matrix are obtained either 
by starting with $\ket{0}^{(+)}$ or with $\ket{0}^{(-)}$. Let us first begin with $\ket{0}^{(+)}$
whose corresponding eigenvalue
$\Lambda_{0}^{(+)}(\lambda)$ is clearly
\EQ
\Lambda_{0}^{(+)}(\lambda)=g^{(+)}[a(\lambda)]^{2}+g^{(-)}[b(\lambda)]^{2}
\EN

The next step is to solve the Bethe ansatz equations 
for the one-particle state $\widetilde{B}(\lambda_{1})\ket{0}^{(+)}$. As a result  we find two possible rapidities
given by
\EQ
\lambda_{1}^{\pm}=-\frac{1}{2}\frac{\left( \sqrt{g^{(+)}} \pm \sqrt{g^{(-)}}\right)}{\left( \sqrt{g^{(+)}} 
\mp \sqrt{g^{(-)}}\right)}
\EN
giving us the following one-particle $\Lambda_{1}^{\pm}$ eigenvalues
\EQ
\Lambda_{1}^{\pm}=a(\lambda)b(\lambda)\left( g^{(+)} + g^{(-)} \right) \pm \sqrt{g^{(+)} g^{(-)}}
\EN

Repeating similar exercise for the 
two-particle state $\widetilde{B}(\lambda_{1})\widetilde{B}(\lambda_{2})\ket{0}^{(+)}$ we have
\EQ
\lambda_{1,2}=\frac{g^{(+)}+g^{(-)}\pm 2 I \sqrt{g^{(+)}g^{(-)}}}{2\left( g^{(-)} - g^{(+)}\right)}
\EN
and the corresponding eigenvalue is
\EQ
\Lambda_{2}(\lambda)=g^{(-)}[a(\lambda)]^{2} + g^{(+)}[b(\lambda)]^{2}
\label{A5}
\EN

An exact diagonalization of the transfer matrix (\ref{transferT}) corroborates 
these four possible eigenvalues for $L=2$. Note
also that (\ref{A5}) is exactly the eigenvalue associated to reference state $\ket{0}^{(-)}$. The others 
eigenvalues (A.1) and (A.3) are easily obtained
from $\ket{0}^{-}$ by noticing that to each solution ${\lambda_{i}^{(+)}}$ 
one can find the corresponding ${\lambda_{i}^{(-)}}$  through the reflection
${\lambda_{i}^{(-)}=-{\lambda_{i}^{(+)}}}$ symmetry.  We have also investigated numerically 
this problem for $L=3,4$ and found that both references states can lead to the complete spectrum of $T(\lambda)$.

\end{document}